# RESONANT ACOUSTIC WAVE ASSISTED SPIN-TRANSFER-TORQUE SWITCHING OF NANOMAGNETS

Austin Roe[1], Dhritiman Bhattacharya[1], and Jayasimha Atulasimha[1,2] *

[1]*Department of Mechanical and Nuclear Engineering, Virginia Commonwealth University, Richmond, VA 23284, USA*
[2]*Department of Electrical and Computer Engineering, Virginia Commonwealth University, Richmond, VA 23284, USA*
* Corresponding author: jatulasimha@vcu.edu

We report the possibility of achieving an order of magnitude reduction in the energy dissipation needed to write bits in perpendicular magnetic tunnel junctions (p-MTJs) by simulating the magnetization dynamics under a combination of resonant surface acoustic waves (r-SAW) and spin-transfer-torque (STT). The magnetization dynamics were simulated using the Landau-Lifshitz-Gilbert equation under macrospin assumption with the inclusion of thermal noise. The resonant magnetization dynamics in the magnetostrictive nanomagnet build over few 10s of cycles of SAW application that drives the magnetization to precess in a cone with a deflection of ~45º from the perpendicular direction. This reduces the STT current density required to switch the magnetization direction without increasing the STT application time or degrading the switching probability in the presence of room temperature thermal noise. This could lead to a pathway to achieve energy efficient switching of spin transfer torque random access memory (STTRAM) whose lateral dimensions can be scaled aggressively despite using materials with low magnetostriction by employing resonant excitation.

The quest to overcome the limitations of random access memory (RAM) to achieve energy efficient and higher speed computing architectures has motivated research to replace volatile CMOS memory devices. One alternative is the use of nanoscale magnetic tunnel junctions (MTJ) that implement non-volatile memory. The magnetization orientation of a MTJ's soft magnetic layer can be switched using different methods [1–7]. One such method utilizes spin transfer torque (STT) which involves applying a current through a hard magnetic layer to polarize electrons that subsequently exert a torque on the magnetization of the soft layer to switch its direction. However, these STT based MTJ devices require ~100 fJ/bit [8] to switch the magnetization which is 1000 times more than the ~100 aJ energy required to switch CMOS devices [9]. This inefficiency could prevent the widespread adoption of pure current driven spin transfer torque (STT) switching [2,10,11]. Another newer technology that addresses some of the shortcomings of STT is spin orbit torque (SOT) induced switching [4,12,13]. However, it results in a 3-terminal memory device that could impede aggressive scaling.

Strain generated via surface acoustic waves (SAW) can also be used to write a bit in an MTJ [14] by controlling the magnetization of its soft magnetostrictive layer. SAW can be created from an interdigitated transducer (IDT) fabricated on a piezoelectric substrate which typically produces Rayleigh (transverse) waves. Rotating the magnetization through the use of SAW is very energy efficient, however, only a ~90° rotation is possible unless dipole coupling [15] or sequential stress along multiple directions is used [16].

Mixed mode SAW and STT is a potential alternative to overcoming the large write energy requirement of STT and complexity of SAW devices. Previous efforts have explored this concept in nanomagnets with in-plane magnetization using low frequency SAW [17], which allows for the quasistatic rotation of the magnetization to an approximately known deflection. STT can be applied to achieve switching once this maximum deflection is reached. However, this approach requires a much higher magnitude SAW, especially in scaled nanomagnets. This is because the stress levels required to produce a large deflection of the magnetization increases with decreased volume of the nanomagnets as the uniaxial magnetic anisotropy $K_u$ increases to ensure an energy barrier ~1 eV between the "0" and "1" states ($K_u\Omega \approx 1$ eV, where $\Omega$ is the volume). Due to this, mixed mode switching has potential issues scaling below 100 nm diameter, even if moderately magnetostrictive materials such as FeGa [18] are used with low Gilbert damping [19]. High magnetostrictive materials such as Terfenol-D [20] will not necessarily achieve a larger magnetization deflection with low stress levels due to the bidirectional coupling between magnetization and strain [21]. This further motivates our resonance approach to overcome the limitations associated with quasistatic SAW excitation, allowing for competitive scalability to smaller lateral dimensions.

We study a hybrid resonant SAW and STT scheme to switch the magnetization of nanomagnets for both in-plane and perpendicular-to-plane magnetization. Fig. 1 demonstrates how these devices can be realized. The SAW is applied over an entire array of nanomagnets and thus adds very little to the energy dissipation per bit switched. Once the nanomagnets have reached maximum deflection (Fig. 1(b) for in-plane, and (c) for out-of-plane), spin-transfer-torque can be applied to minimize the spin current required to switch as compared to a non-stressed state.

Modelling of the magnetization dynamics was performed by solving the Landau-Lifshitz-Gilbert (LLG) equation [22] with inclusion of damping like spin transfer torque term [23]:

$$(1 + \alpha^2)\frac{d\vec{M}}{dt} = \gamma \vec{M} \times \vec{H}_{eff} - \frac{\alpha\gamma}{M_s}[\vec{M} \times (\vec{M} \times \vec{H}_{eff})] + \frac{\alpha\gamma}{M_s}\beta\varepsilon(M \times (M_p \times M)) \quad (1)$$

$$\beta = \left|\frac{\hbar}{\mu_0 e}\right|\frac{J}{l_{th}M_s} \quad (2)$$

where M is the is the magnetization, γ is the gyromagnetic ratio, α is the Gilbert damping coefficient, $M_s$ is the saturation magnetization and $M_p$ is the magnetization in the direction of the STT polarization, $\hbar$ is the Planck constant, $\mu_0$ is the permeability of free space, e is the charge of an electron, J is the current density of the STT, and $l_{th}$ is the thickness of the nanomagnets. The effective field was calculated from the total energy of the system:

$$\vec{H}_{eff} = -\frac{1}{\mu_0 \Omega}\frac{dE}{d\vec{M}} \quad (3)$$

$$E = E_{stress\ anisotropy} + E_{shape\ anisotropy} + E_{PMA} \quad (4)$$

$$E_{shape\ anisotropy} = \left(\frac{\mu_0}{2}\right)\Omega[N_{d\_xx}M_x^2 + N_{d\_yy}M_y^2 + N_{d\_zz}M_z^2] \quad (5)$$

$$E_{stress\ anisotropy} = -\frac{3}{2}[\lambda_s \sigma \Omega]\sin^2\theta_i \sin^2\phi_i \quad (6)$$

$$E_{PMA} = K_{s0}\cos^2\theta_i \quad (7)$$

with Ω representing the volume of the nanomagnets, $N_{d\_xx}$, $N_{d\_yy}$, and $N_{d\_zz}$ the demagnetization factors in the respective directions, $\lambda_s$ the saturation magnetostriction, σ the stress produced by the SAW on the nanomagnets, $\theta_i$ and $\phi_i$ the polar and azimuthal angles, respectively, of the magnetization, and $K_{s0}$ the surface anisotropy constant. We note the effective field due to stress is calculated from uniaxial stress only (cyclic tension and compression) due to the SAW. The stress in the in-plane direction orthogonal to SAW propogation, which sees the opposite stress (cyclic compression and tension) due to Poisson's effect, is neglected. This does not change the magnetization dynamics qualitatively, and only makes the stress amplitude we estimate conservative (a smaller stress will produce the same effective field if stress in the other direcition is accounted for).

Table I lists the values of the material properties of $Fe_{81}Ga_{19}$ used in these simulations. It was chosen as it has moderate magnetostriction and low Gilbert damping [18,19].

*Table I: FeGa material properties*

| Parameters | $Fe_{0.81}Ga_{0.19}$ |
|---|---|
| **Saturation Magnetization ($M_s$)** | $0.8 \times 10^6$ A/m |
| **Gilbert Damping (α)** | 0.015 |
| **Gyromagnetic Ratio (γ)** | $2.2 \times 10^5$ |
| **Magnetostriction ($\lambda_s$)** | 350 ppm |

The magnetization dynamics were simulated as follows. The resonant SAW was applied for several nanoseconds to build the maximum deflection of the magnetization. The magnitude of the SAW was chosen so that no switching from only the SAW excitation occurs. When the maximum deflection was reached, the minimum STT current density needed to maintain low switching error probability (for example, 99.9% switching in case of out-of-plane switching limited by the number of computations we could perform) was applied. The time of removal of the SAW (as long as this was after STT application was completed) made no difference to the final switching probability.

**IN-PLANE SWITCHING WITH SAW + STT:** Simulation of elliptical nanomagnetic disks with in-plane magnetic anisotropy is shown in Fig 1(b) with dimensions of 40 x 30 nm along the easy and hard axis, respectively, and a thickness of 6 nm. SAW was applied to the nanomagnets and the resonance frequency, which creates the maximum deflection from the magnetic easy axis at the given SAW magnitude, was found. The SAW frequency was found to be double that of the frequency at which the magnetization precesses about the easy axis. When the magnet is compressed along its long (easy) axis, the magnetization rotates towards the magnetic hard axis on one side, and when a tensile stress is applied, the magnetization returns to the easy axis. This process isrepeated with magnetization moving towards the opposite hard axis in the next cycle, leading to a doubling of the frequency for the applied SAW. If resonance is achieved, the deflection of the magnetization from the easy axis will increase until a maximum deflection is reached over a few 10s of cycles. Without thermal noise (Fig 2(a)), the STT current density required ($15 \times 10^{11}$ A/m² with a SAW of 23 MPa at 10.7 GHz) is lowered drastically compared to when no SAW is applied. To achieve switching in the same time without the application of SAW, a current density ~3 times larger was needed. This translates to nearly one order of magnitude lower energy (3 times current is 9 times write energy, assuming similar application time). However, in the in-plane case, this potential energy saving is severely negated by thermal noise effects as discussed next.

**THERMAL NOISE:** The magnitude of the STT current density needed to achieve 99% switching of the magnetization from the positive to the negative y-axis in the presence of thermal noise and applied SAW of 10 MPa at 9.42 GHz was $10 \times 10^{11}$ A/m$^2$, shown in Fig. 2(c). Now when the SAW was removed the same current density magnitude sufficed to achieve 99% switching, indicating that the SAW was not helping lower the STT write current for the in-plane switching in the presence of thermal noise (Fig. 2 b). This can be explained as follows.

Prior to STT application, a large difference can be seen between the magnetization deflection from the thermal noise without SAW to that with thermal noise with SAW. This indicates that the SAW indeed produced higher deflection even with the thermal noise. However, once STT of sufficient magnitude is applied, there is no difference in switching probability between the no-SAW and SAW case. In previous calculations without thermal noise, it was possible to determine apriori the specific timing to apply the STT so that the magnetization is at a point of maximum deflection. However, due to thermal fluctuations, while there is higher deflection with SAW, the inability to apply STT at the maximum deflection negated the benefit of any added deflection from the SAW and no difference was seen between the switching probability in the two cases, as shown in Fig. 2(b) and 2(c).

**OUT-OF-PLANE SWITCHING WITH SAW + STT:** Simulation of the magnetization dynamics of cylindrical disks with a diameter of 50 nm and thickness 1.5 nm with perpendicular magnetic anisotropy is shown in Fig. 1 (c). Initially, the magnetization points in the out-of-plane direction (+z-axis), and a continuous SAW was applied across the magnets to find the resonant frequency, which is dependent on the anisotropy of the nanomagnets and magnitude of the SAW (as the oscillation is non-linear). This SAW applies compression and tension along the y-axis of the nanomagnet (and vice versa along the x-axis). This cyclical stressing on the magnet rotates the magnetization from pointing directly out-of-plane to begin to precess about the z-axis, as shown in Fig. 1(c). As a result of the resonant SAW, the magnetization precesses further from the z-axis and the cone of rotation becomes larger as the magnetization approaches the x-y plane of the nanomagnet.

Once the maximum deflection of the magnetization from the out-of-plane direction was achieved, spin-transfer-torque is applied. As in the previous case, once the magnetization is switched and the STT current is withdrawn, the SAW can be concurrently removed or continued to run for a few cycles before withdrawal as it is not enough to switch the magnetization on its own.

Previously with the in-plane calculations, the inability to selectively apply the STT when the deflection was at a maximum was an issue. However, the manner in which the magnetization precesses around the out-of-plane axis eliminates the need to time the application of the spin torque. This is because for a given polar deflection, θ of the magnetization, which is reached and stabilized after a few cycles, the azimuthal angle, Φ shown in Fig. 1 at which the magnetization is oriented when the STT is applied does not affect its efficacy.

Fig. 3(a) shows the effect of SAW on deflection of the magnetization with and without thermal fluctuations. Without any SAW, thermal noise alone deflects the magnetization to a maximum of ~10° from the z-axis. On the other hand, when SAW of 100 MPa at a resonant frequency of 9.95 GHz is applied and the magnetization allowed to build to its maximum deflection, an average deflection of ~45° is seen in both the presence and absence of thermal noise. The thermal noise merely makes the deviation from the mean deflection of ~45° much higher than the case of precession without thermal noise. Fig. 3(b) shows the trajectories of magnetization switching in the presence of thermal noise for applied resonant SAW and STT. At a fixed current density and STT application time the switching probabilities (trajectories that switch) are much higher with SAW than without the SAW.

**OUT-OF-PLANE SWITCHING STT ONLY VS. SAW + STT ENERGY COMPARISON:** At a current density of $1.9 \times 10^{11}$ A/m$^2$ applied for 0.7 ns along with a SAW magnitude of 100 MPa at 9.95 GHz, 99.9% of the simulated nanomagnets successfully switched from the positive to the negative z-axis. Without any SAW, at this same current density and STT application time, roughly 89% of the magnets switched. This error could be reduced to 99.9% in the no SAW case if the current density was doubled to $3.8 \times 10^{11}$ A/m$^2$, as shown in Fig. 3(c). Simulation of SAW at both 30 MPa at 11.36 GHz and 60 MPa at 10.79 GHz was also conducted and a current density of $3 \times 10^{11}$ A/m$^2$ and $2.2 \times 10^{11}$ A/m$^2$, respectively, was required to achieve 99.9% switching probability. Keeping the current density magnitude fixed at $2 \times 10^{11}$ A/m$^2$ and adjusting the STT application time yielded similar results. While previously, 99.9% of nanomagnets switched at 100 MPa and 0.7 ns, the case with no SAW needed double the amount of time (1.4 ns) to ensure the same switching probability.

With required current density being just half that compared to the case without any SAW, the energy savings are four times that of the pure STT case. With materials such as Rare Earth substituted YIG [24] that have low magnetostriction but extremely small damping, at least an order of magnitude energy saving with similar or better error rates could be achieved. This could open the pathway to achieve desired scalability and be competitive with current CMOS memory implementations while having the added advantage of non-volatility.

In summary, we have shown that application of resonant SAW and STT can be more energy efficient in switching nanomagnetic soft layer of p-MTJs than switching with only STT. This process can be further optimized to increase the energy savings and switching time. These theoretical results could stimulate experimental work, ultimately resulting in the development of more energy efficient STTRAM.

A.R., D.B. and J.A are supported National Science Foundation SHF Small grant #CCF-1815033.

**References:**

[1] M. T. Alam, M. J. Siddiq, G. H. Bernstein, M. Niemier, W. Porod, and X. S. Hu, IEEE Trans. Nanotechnol. **9**, 348 (2010).

[2] D. C. Ralph and M. D. Stiles, J. Magn. Magn. Mater. **320**, 1190 (2008).

[3] M. Yamanouchi, D. Chiba, F. Matsukura, and H. Ohno, Nature **428**, 539 (2004).

[4] D. Bhowmik, L. You, and S. Salahuddin, Nat. Nanotechnol. **9**, 59 (2014).

[5] S. Manipatruni, D. E. Nikonov, and I. A. Young, Appl. Phys. Express **7**, 103001 (2014).

[6] N. Tiercelin, Y. Dusch, V. Preobrazhensky, and P. Pernod, J. Appl. Phys. **109**, 07D726 (2011).

[7] K. Roy, S. Bandyopadhyay, and J. Atulasimha, Appl. Phys. Lett. **99**, 063108 (2011).

[8] J. J. Nowak, R. P. Robertazzi, J. Z. Sun, G. Hu, J.-H. Park, J. Lee, A. J. Annunziata, G. P. Lauer, R. Kothandaraman, E. J. O'Sullivan, P. L. Trouilloud, Y. Kim, and D. C. Worledge, IEEE Magn. Lett. **7**, 1 (2016).

[9] S. Datta, V. Q. Diep, and B. Behin-Aein, (2014).

[10] J. C. Slonczewski, J. Magn. Magn. Mater. **159**, L1 (1996).

[11] L. Berger, Phys. Rev. B **54**, 9353 (1996).

[12] L. Liu, C.-F. Pai, Y. Li, H. W. Tseng, D. C. Ralph, and R. A. Buhrman, Science **336**, 555 (2012).

[13] I. Mihai Miron, G. Gaudin, S. Auffret, B. Rodmacq, A. Schuhl, S. Pizzini, J. Vogel, and P. Gambardella, Nat. Mater. **9**, 230 (2010).

[14] N. D'Souza, A. Biswas, H. Ahmad, M. Salehi Fashami, M. Mamun Al-Rashid, V. Sampath, D. Bhattacharya, M. Ahsanul Abeed, J. Atulasimha, and S. Bandyopadhyay, Nanotechnology **29**, 49 (2018).

[15] V. Sampath, N. D'Souza, G. M. Atkinson, S. Bandyopadhyay, and J. Atulasimha, Appl. Phys. Lett. **109**, 102403 (2016).

[16] A. K. Biswas, H. Ahmad, J. Atulasimha, and S. Bandyopadhyay, Nano Lett. **17**, 3478 (2017).

[17] A. K. Biswas, S. Bandyopadhyay, and J. Atulasimha, Appl. Phys. Lett. **103**, 232401 (2013).

[18] A. Clark, M. Wun-Fogle, J. B. Restorff, and T. A. Lograsso, Mater. Trans. **43**, 881 (2002).

[19] D. B. Gopman, V. Sampath, H. Ahmad, S. Bandyopadhyay, and J. Atulasimha, IEEE Trans. Magn. **53**, (2017).

[20] L. Sandlund, M. Fahlander, T. Cedell, A. E. Clark, J. B. Restorff, and M. Wun-Fogle, J. Appl. Phys. **75**, 5656 (1994).

[21] Z. Xiao, R. Lo Conte, C. Chen, C.-Y. Liang, A. Sepulveda, J. Bokor, G. P. Carman, and R. N. Candler, Sci Rep. **8**, 5207 (2018).

[22] T. L. Gilbert, IEEE Transactions on Magnetism, **40** (6) 3443 (2004).

[23] J. Z. Sun, Phys. Rev. B **62**, 570, 2000.

[24] E. R. Rosenberg, L. Beran, C. O. Avci, C. Zeledon, B. Song, C. Gonzalez-Fuentes, J. Mendil, P. Gambardella, M. Veis, C. Garcia, G. S. D. Beach, and C. A. Ross, Phys. Rev. Mater. **2**, 94405 (2018).

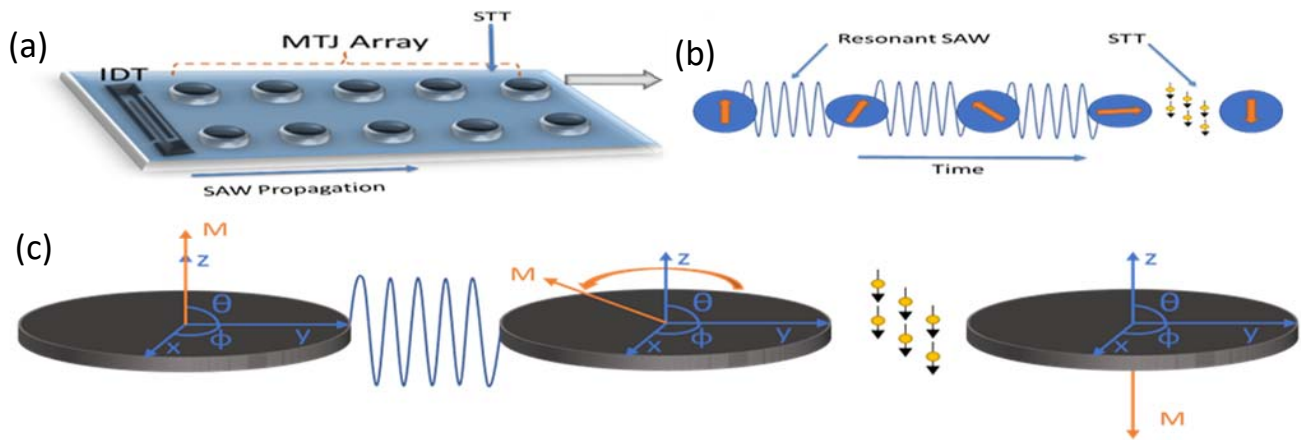

*Figure 1: (a) MTJ array switched with resonant SAW and STT (b) Magnetization dynamics with resonant SAW + STT switching of in-plane magnetization (c) Magnetization dynamics with resonant SAW + STT switching of out-of-plane magnetization.*

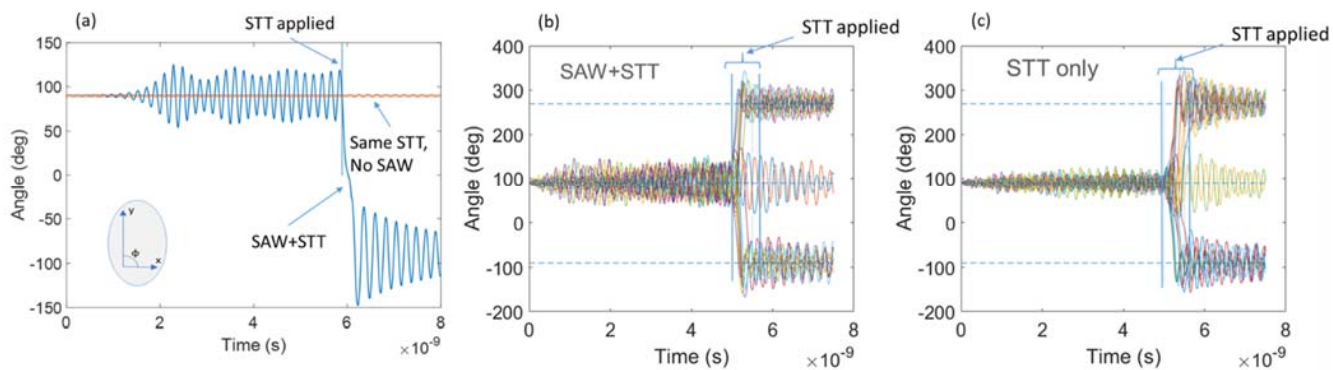

*Figure 2: In-plane magnetization dynamics simulations with (a) no thermal noise, (b) resonant STT and SAW, and (c) STT only and no SAW. NOTE: Angle shown is the azimuthal angle φ with ~90° degree being the initial orientation and both ~270° and ~ -90° representing a successful switch, while return to ~90° degree being an unsuccessful switching event.*

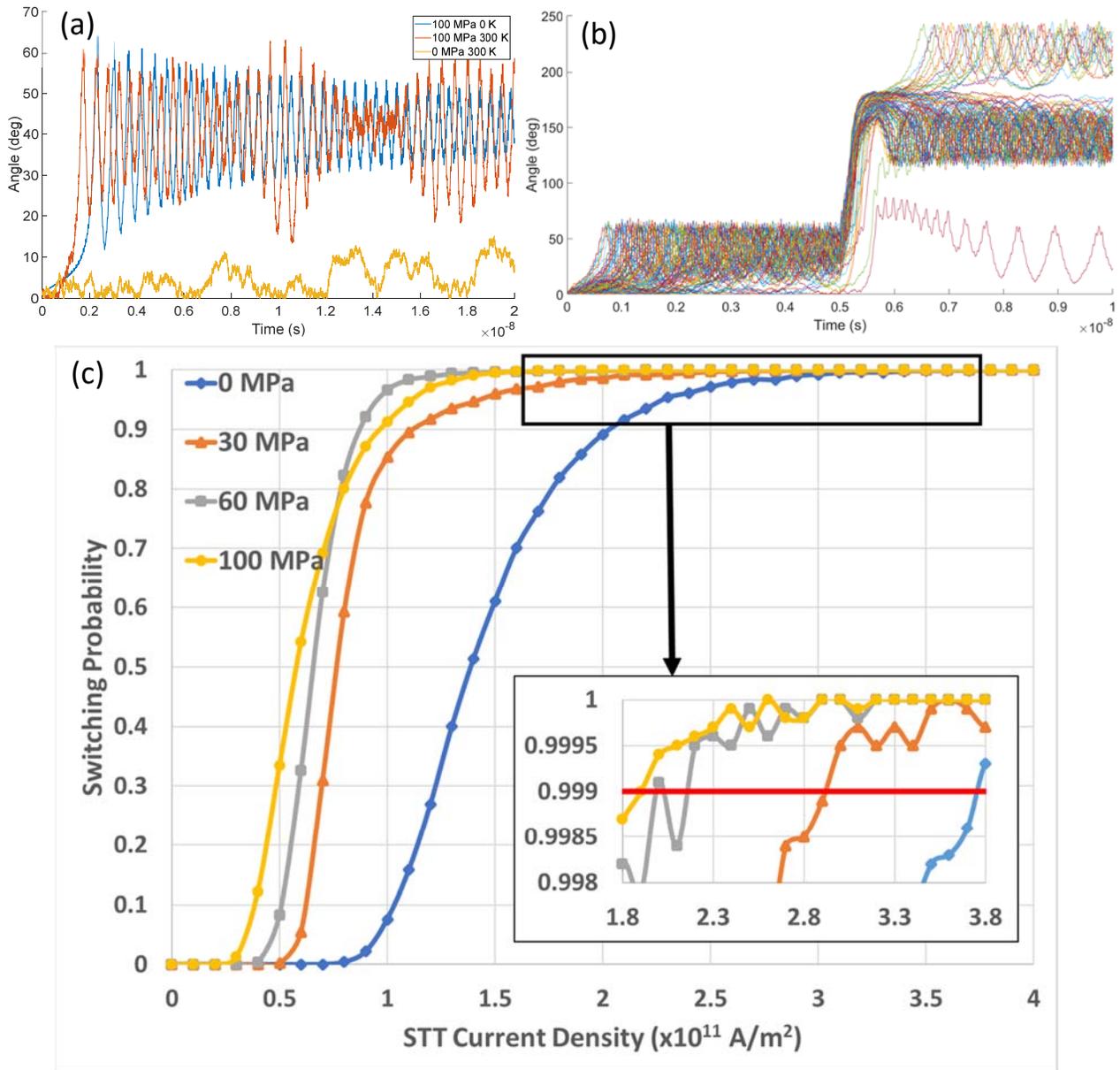

*Figure 3: Out-of-plane magnetization dynamics simulations with (a) comparison of resonant SAW with and without thermal noise to purely thermal noise and (b) switching of out-of-plane magnetization with resonant SAW and STT (angle shown is the polar angle, θ) (c) Switching probability vs. STT current density for three different SAW magnitudes as well as for no SAW appled.*